\documentclass{article}

\usepackage{graphicx}
\usepackage{amssymb}
\usepackage{amsmath}
\usepackage{lineno}
\usepackage{color}
\usepackage{hyperref}
\usepackage{multirow}

\DeclareMathOperator*{\argmax}{argmax}

\begin{document}


\title{Understanding diseases as increased heterogeneity: a complex network computational framework}

\author{Massimiliano Zanin \and Juan Manuel Tu\~nas \and Ernestina Menasalvas}
\date{%
    Centro de Tecnolog\'ia Biom\'edica, Universidad Polit\'ecnica de Madrid, 28223 Madrid, Spain
    \\ \today
}

\maketitle

\begin{abstract}
Due to the complexity of the human body, most diseases present a high inter-personal variability in the way they manifest, {\it i.e.} in their phenotype, which has important clinical repercussions - as for instance the difficulty in defining objective diagnostic rules. We here explore the hypothesis that signs and symptoms used to define a disease should be understood in terms of the dispersion (as opposed to the average) of physical observables. To that end, we propose a computational framework, based on complex networks theory, to map groups of subjects to a network structure, based on their pairwise phenotypical similarity. We demonstrate that the resulting structure can be used to improve the performance of classification algorithms, especially in the case of a limited number of instances, both with synthetic and real data sets. Beyond providing an alternative conceptual understanding of diseases, the proposed framework could be of special relevance in the growing field of personalised, or $N$-to-$1$, medicine.

\vspace{0.5cm}
\end{abstract}


\section{Introduction}

This decade is witnessing important improvements in the way many diseases and disorders are understood and treated, thanks to advancements spanning from genetics to big data analysis; yet, a large share of them is still eluding our comprehension. Such gap is partly the result of the complexity of translating the symptoms and signs observed in a condition, to the real causes underpinning it - an inverse problem equivalent to the ``genotype to phenotype'' one \cite{benfey2008genotype}. Still an additional important element has mostly been neglected: the way such symptoms and signs are defined.

A standard and {\it prima facie} good option entails detecting changes in the expected value (that is, in the mean) of some variables - {\it i.e.} fever is considered a sign of viral infection as the body temperature is usually higher in patients than in control subjects.
On the other hand, an alternative approach recognises that the complexity of the human body implies that multiple elements may interact in different ways depending on the characteristics of the patient. In other words, if diseases are to be understood as network perturbations \cite{del2010diseases}, such perturbations can affect an individual node in different ways; and, as a result, the observable that increases in one subject may decrease in a different person.
According to this approach to diseases, control subjects are such because they maintain the standard homoeostasis and allostasis \cite{schulkin2004allostasis}, and are hence homogeneous; and, on the other hand, patients loose such homogeneity, and are thus characterised by a higher variability.
This inter-personal variability has recently been recognised within the personalized medicine paradigm \cite{schork2015personalized}; but its origins can be traced back to Hippocrates himself \cite{offit2011personalized}.
In the light of this idea, a different approach of understanding symptoms may be more appropriate, {\it i.e.} one in which the average value of an observable is not as relevant as the corresponding variability. Diseases could then be characterised by abnormal values, {\it i.e.} by an increase in the standard deviation of the corresponding distribution, in sets of parameters that may differ from individual to individual.  

A simple example can help clarifying this concept. Suppose a set of control subjects, all characterised by two physiological parameters $a$ and $b$ described by a truncated normal distribution $| \mathcal{N}(0, 1) |$. Most control subjects (actually the $95.45\%$ of them) will thus have values for $a$ and $b$ in the range $[0, 2]$.
Let us further suppose an additional population of people suffering from a disease, whose effect is to unbalance the bodily homoeostasis; as a consequence, patients may be described by the distribution $a_p \sim b_p \sim | \mathcal{N}(0, 2) |$ - note the increased standard deviation. What would a classical symptom analysis yield? The answer would be a fuzzy definition: some patients will show no substantial modification, being both $a$ and $b$ within the standard range; others will have abnormal high values of $a$ or of $b$; and a few of both $a$ and $b$. The pathology would thus be difficult to explain, and even more difficult to diagnose.
The alternative approach entails describing the disease in terms of the variability observed in each group. Control subjects will thus be described as conforming an homogeneous group, with $E[ |(a, b)| ] = 2 / \sqrt{\pi}$; patients a more heterogeneous one, with $E[ |(a_p, b_p)| ] = 4 / \sqrt{\pi}$.

If the hypothesis here presented is true, such that diseases should be defined in terms of higher variability, the expected result would be a high degree of {\it fuzziness} in standard disease definitions, with expressions like ``symptoms may include'' - as such symptom would appear for some patients, but not in all of them, as depicted in the previous example.
This is indeed found, for instance, in the large heterogeneity observed in cancer: cells belonging to a same cancer type, or even to the same patient, can display a strong phenotypic and functional variability \cite{felipe2013cancer, meacham2013tumor}. 
Such effect is not exclusive of genetics, but appear in other medical fields. Psychology and psychiatry yields several relevant examples: {\it e.g.} the obsessive-compulsive disorder, displaying an univocal definition and yet a wide range of phenotypes \cite{bloch2008meta, martoni2015evaluating}; similarly, depression \cite{kok2012can}; attention-deficit/hyperactivity disorder \cite{waahlstedt2009heterogeneity}; or borderline personality disorder \cite{bornovalova2010understanding}. As the study of the (possibly) most complex system of all, this problem is also faced in neuroscience, {\it e.g.} in Huntington's \cite{rosas2008cerebral}, Parkinson's \cite{kehagia2010neuropsychological}, or Alzheimer's \cite{lambert2007genetic} diseases, all of them eluding a simple characterization.
This has important clinical repercussions, as it is difficult to provide practitioners with a set of objective diagnostic rules, not partly relying on personal experience and judgement \cite{kienle2011clinical}.

In this contribution we explore the idea of a paradigm shift in diseases characterisation, based on moving from the {\it average} to the {\it dispersion} in the analysis of signs.
For this we propose a computational framework, based on complex networks theory \cite{strogatz2001exploring, newman2003structure}, for the characterisations of groups of healthy subjects and patients. It is based on the identification of the two sets of features that optimise the homogeneity of respectively control subjects and patients; and on the creation of a graph structure representing the pairwise distance between individuals according to those sets. Besides yielding a compact representation of the disease under study, we show how this framework can be successfully used to improve the outcome of classification ({\it i.e.} diagnosis) tasks, especially in situations with limited number of training instances. Results are shown for synthetic and real data sets, the latter representing a wide range of biomedical problems.

\section{Methods}

\subsection{Convergence/divergence network creation}
\label{sec:creation}

Based on the previous considerations, we here describe a computational methodology for creating a network structure in which instances belonging to previously labelled groups are organised according to their internal affinity. 
Without loss of generality, we here consider the case of two groups, respectively $g_1$ and $g_2$, such that two networks naturally emerge: one ($N_1$) maximising the coherence of instances belonging to $g_1$, and a second one (namely $N_2$) maximising the coherence of $g_2$.
For the sake of clarity, we here name the two structures respectively as {\it convergence} and {\it divergence} networks - as, in the former, $g_1$ instances converge towards a common pattern, while in the latter they diverge from it.
While the proposed methodology can naturally be extended to additional groups and networks, this example has been chosen due to its simplicity and relevance - as, for instance, most biomedical classification problems deal with characterising control subjects {\it vs.} patients. 

In the simplest situation, each instance $i$ would be characterised by a set of features $f_i$. A distance matrix $D$ is then defined, whose element $d_{i, j}$ corresponds to the cosine distance between the two instances $i$ and $j$:

\begin{equation}
d_{i,j} = D_C(f_i, f_j) = 1 - \frac{f_i \cdot f_j}{||f_i|| \cdot ||f_j||}.
\end{equation}

Note that other distance metrics can be used, as will be discussed in Sec. \ref{sec:val:synth}. The internal coherence of each group is then estimated through the average of all pairwise internal distances, {\it i.e.}:

\begin{eqnarray}
c_1 = \frac{1}{n_1 ( n_1 - 1 )} \sum _{i \in g_1, j \in g_1, j \neq i} d_{i, j}, \\
c_2 = \frac{1}{n_2 ( n_2 - 1 )} \sum _{i \in g_2, j \in g_2, j \neq i} d_{i, j},
\end{eqnarray}

$n_1$ and $n_2$ respectively representing the number of instances in groups $g_1$ and $g_2$.

While in this situation, {\it i.e.} using all features, instances of the two groups may naturally be clustered, in general one is interested in finding the two subsets of features, $f^c$ and $f^d$, optimising the creation of the two networks. This corresponds to the following maximisation problem:

\begin{eqnarray}
\argmax_{f^c} \log \frac{c_2}{c_1} \bigg\rvert _{f^c}, \\
\argmax_{f^d} \log \frac{c_1}{c_2} \bigg\rvert _{f^d}.
\end{eqnarray}

Note that, in the first case, we are creating the {\it convergence} network, {\it i.e.} the network in which the first set of instances are more homogeneous, and we thus need to maximise $c_2 / c_1$. On the other hand, the {\it divergence} network requires the instances of the first group to be more heterogeneous than the second group, {\it i.e.} the maximisation of $c_1 / c_2$.

\begin{figure}[!tb]
\begin{center}
\includegraphics[width=0.99\textwidth]{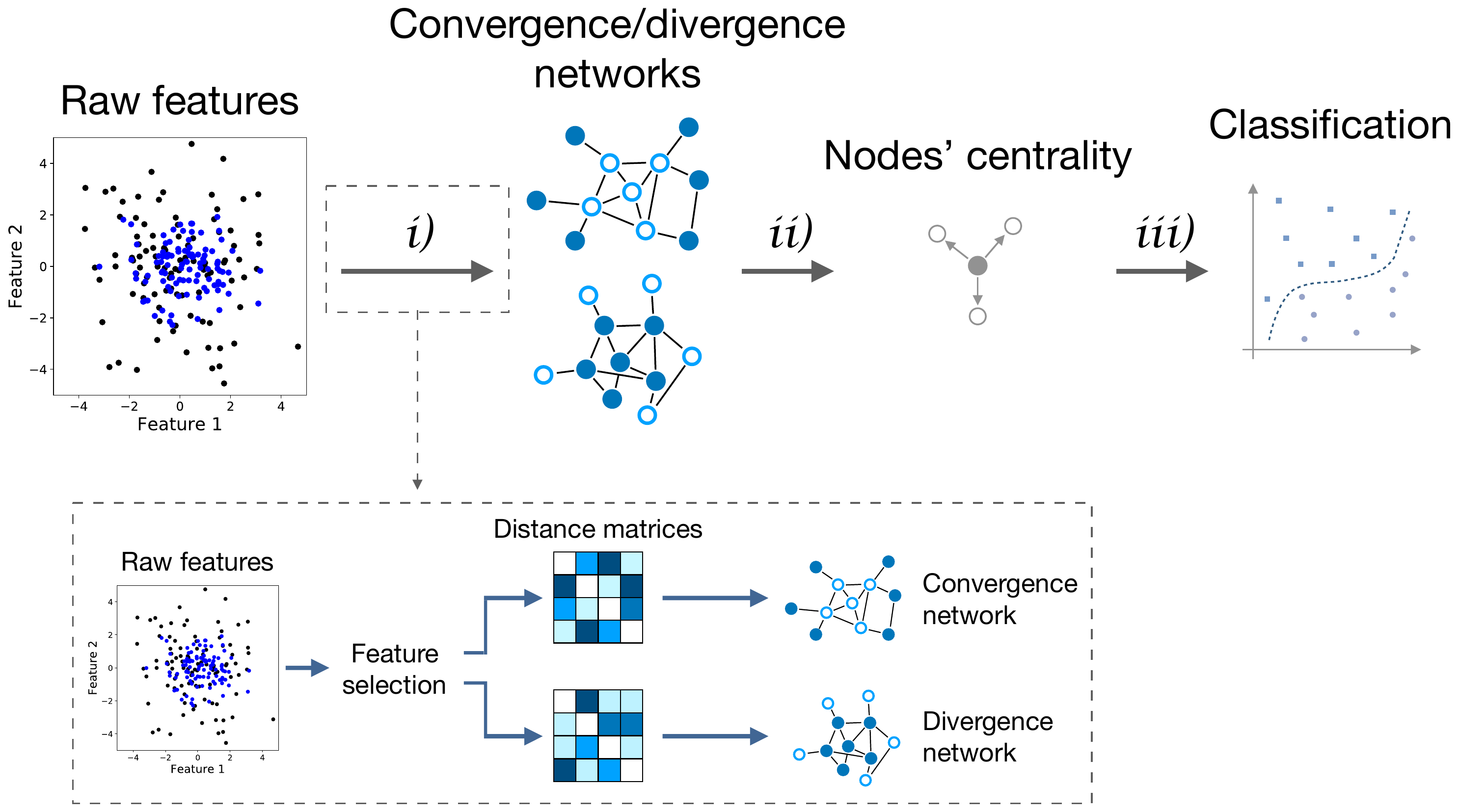}
\caption{Graphical representation of the proposed methodology. The lower inset depicts the process of creating the two {\it convergence}/{\it divergence} networks, while the upper panel integrates them into a general classification task. See main text for definitions and details.}\label{fig:Method}
\end{center}
\end{figure}

As depicted in Fig. \ref{fig:Method}, bottom part, the result of this first phase is a set of two fully connected networks (cliques), respectively called {\it convergence} and {\it divergence}.
In both networks, nodes represent instances (irrespectively of their group); and the weight of links between them the distance, in the selected feature sub-space, between the corresponding instances. For the sake of clarity, node colours in Fig. \ref{fig:Method} depict the group affiliation, with each group laying in the centre of the corresponding network.

\subsection{From networks to classification}

The previously obtained convergence / divergence networks have an intrinsic value, as they can be used to represent the structures created by patients' similarity, and to identify the most relevant features in a data set.
Nevertheless, if the objective is to perform a classification task ({\it e.g.} a diagnosis), it is necessary to transform the two networks into a set of features that can be understood by a data mining algorithm - as directly training a classification algorithm with the adjacency matrices is usually not efficient \cite{zanin2016combining}. As the instances of the first group are, by construction, more connected in the convergence network (as are the instances of the second group in the divergence one), it seems natural to use a centrality measure to describe each node. Specifically, we consider three standard centralities: ({\it i}) the {\it closeness} centrality, defined as the inverse of the average distance from the target node to any other node in the network \cite{okamoto2008ranking}; ({\it ii}) the {\it betweenness} centrality, defined as proportional to the number of shortest paths passing through the considered node \cite{barthelemy2004betweenness}; and ({\it iii}) the {\it eigenvector} centrality, a centrality measure that assess the importance of a node as a function of the importance of its neighbours \cite{bonacich2007some}.
Note that, while all the three metrics assess the importance of nodes from a propagation perspective, they tackle the problem from different points of view.
Consequently, there is no {\it a priori} way of predicting which one of them is more appropriate for the problem here considered, and in what follows all three will be independently tested.

In the proposed framework, each instance of the original data set is finally characterised by two features, {\it i.e.} its centrality in the convergence and divergence networks, as depicted in Fig. \ref{fig:Method}. It is worth noting that this represents an important reduction in the dimensionality of the problem, from $|f|$ to $2$. As is well known in data mining \cite{guyon2003introduction}, such drastic reduction in the number of features yields important benefits, as a lower risk of overfitting, the possibility of working with less instances, and an improved computational cost. If this usually comes at the cost of loosing part of the encoded information, in subsequent sections we will show how here this is not a concern, as enough information is encoded in the networks to actually allow to improve the outcome of several classification tasks.

\section{Validation with synthetic data sets}
\label{sec:val:synth}

In order to evaluate the added value yielded by the proposed framework in a classification task, we here report results corresponding to an ensemble of synthetic random data sets. These have been designed to be of maximal simplicity, while enabling a sensitivity analysis of the algorithm with respect to their characteristics.

First of all, we suppose a binary classification problem, {\it e.g.} aimed at discriminating between healthy people and patients suffering from a given disease. A total of $n_i$ instances are considered, equally distributed between the two classes - in order to avoid classification biases.
Each instance is then described by a set of $n_t$ features, whose values are drawn from the normal distributions defined in Tab. \ref{tab:SynthFeatures}.
These are designed to make some features have a higher variance in patients; some a higher variance in control subjects; and finally, to make some features irrelevant, thus only encoding noise.
Three parameters control the shape of the distributions, namely:

\begin{itemize}
\item $\sigma_2$, increased standard deviation of the relevant features. This parameter is used to simulate the increased heterogeneity in control subjects and patients, hence $\sigma_2 > 1$.

\item $n_r$, number of relevant features for each class, {\it i.e.} features whose heterogeneity is higher than for the other class.

\item $n_t$, total number of features in the data set. A given number of features, namely $n_t - 2n_r$ are completely random and do not depend on the instance's class. Note that $n_t \geq 2n_r$.
\end{itemize}

\begin{table*}[!tb]
\centering
\begin{tabular}{|l|c|c|c|}
\hline
Instance class & Features $[0, n_r)$ & Features $[n_r, 2n_r)$ & Features $[2n_r, n_t)$ \\ \hline
Class 1 & $\mathcal{N}(0, 1)$ & $\mathcal{N}(0, \sigma_2)$ & $\mathcal{N}(0, 1)$ \\ \hline
Class 2 & $\mathcal{N}(0, \sigma_2)$ & $\mathcal{N}(0, 1)$ & $\mathcal{N}(0, 1)$ \\ \hline
\end{tabular}
\caption{Characteristics of the distributions from which synthetic data are drawn. See main text for details.}
\label{tab:SynthFeatures}
\end{table*}

\begin{figure}[!tb]
\begin{center}
\includegraphics[width=0.99\textwidth]{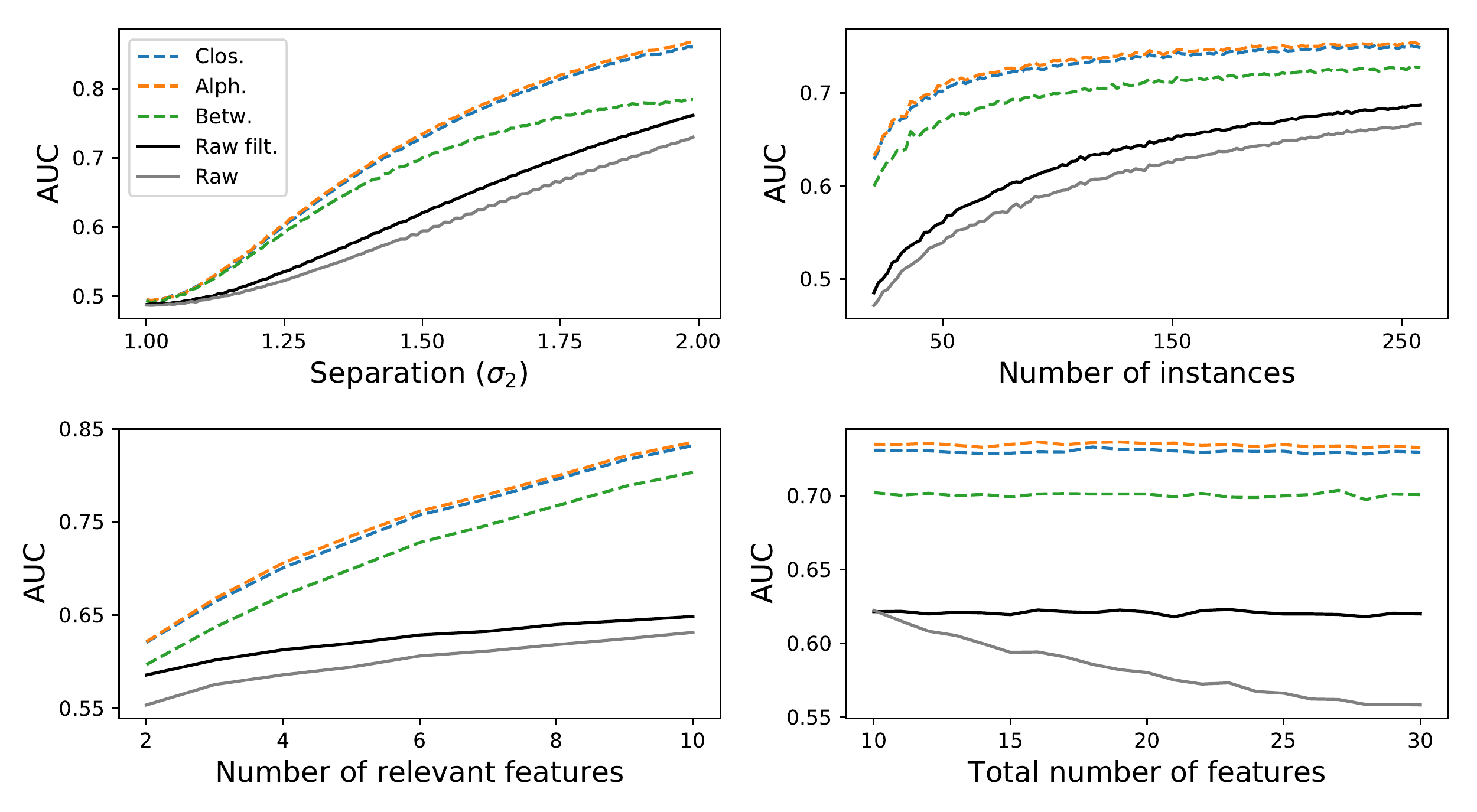}
\caption{Classification score in synthetic data sets, using the raw data (black and grey solid lines), and the new features extracted from the convergence/divergence networks (blue, orange and green dashed lines, respectively for closeness, alpha and betweenness centralities). From left to right, top to bottom, the four panels depict the score as a function of the increased variability of data ($\sigma_2$), the number of instances in the data set, the number of relevant features ($n_r$), and the total number of features ($n_t$). Results correspond to the average of $10^4$ realisations.}\label{fig:ResSynth}
\end{center}
\end{figure}

These synthetic data sets have then been used to perform a classification task, using Random Forests (RFs) \cite{breiman2001random} and a Leave-One-Out Cross Validation (LOOCV) \cite{efron1983leisurely}. Results, in terms of the AUC of the classification \cite{fawcett2006introduction} and as a function of the four data set parameters, are shown in Fig. \ref{fig:ResSynth}. The five curves respectively depict:

\begin{itemize}
\item (Grey solid line) The score obtained by training the RFs using the raw data.
\item (Black solid line) The score obtained by training the RFs using the first $2n_r$ features of the raw data, {\it i.e.} the features that actually encode information.
\item (Blue/orange/green dashed lines) The score obtained by using the convergence/divergence network centrality data - respectively closeness, alpha and betweenness centralities.
\end{itemize}

Some interesting conclusions can be drawn. First of all, the proposed approach allows to get a classification score consistently higher than the one obtained by only using the raw data. While this may be a consequence of using a tailored data set, in Sec. \ref{sec:val:real} we will see how this also happens in several real data sets. The convergence / divergence networks thus seem to be an efficient way of synthesising biological information.

Secondly, it can be appreciated that all AUCs increase with the separation ($\sigma_2$), the number of instances ($n_i$) and the number of relevant features ($n_r$), as more information is encoded in the data set. A different behaviour is observed when the total number of features ($n_t$) is increased, while keeping $n_r$ constant: while the network performance is constant, the classification with all raw features degrades. This suggests that the proposed methodology is not affected by redundant and useless information, as the networks are inherently performing a feature selection.

Additionally, convergence/divergence networks yield a relatively higher AUC with low separations ($\sigma_2$), low number of instances ($n_i$) and high number of relevant features ($n_r$). As may be expected, RFs are not efficient in extracting information that is sparsely encoded over a large number of features, when differences between classes are minimal, and when few instances are available for training. In all these cases, the networks help making explicit the information that may not directly be accessible to the classification algorithm.

Finally, in all cases the alpha centrality is the network metric yielding the best results, followed by the closeness and the betweenness centralities. Possibly this is due to the alpha centrality's higher capacity for accounting for the global (macro-scale) structure of the network.

For the sake of completeness, three aspects of the analyses reported in Fig. \ref{fig:ResSynth} require further consideration: the classification algorithm, the shape of the probability distribution underlying the data, and the way distances between subjects are measures. 

\begin{figure}[!tb]
\begin{center}
\includegraphics[width=0.99\textwidth]{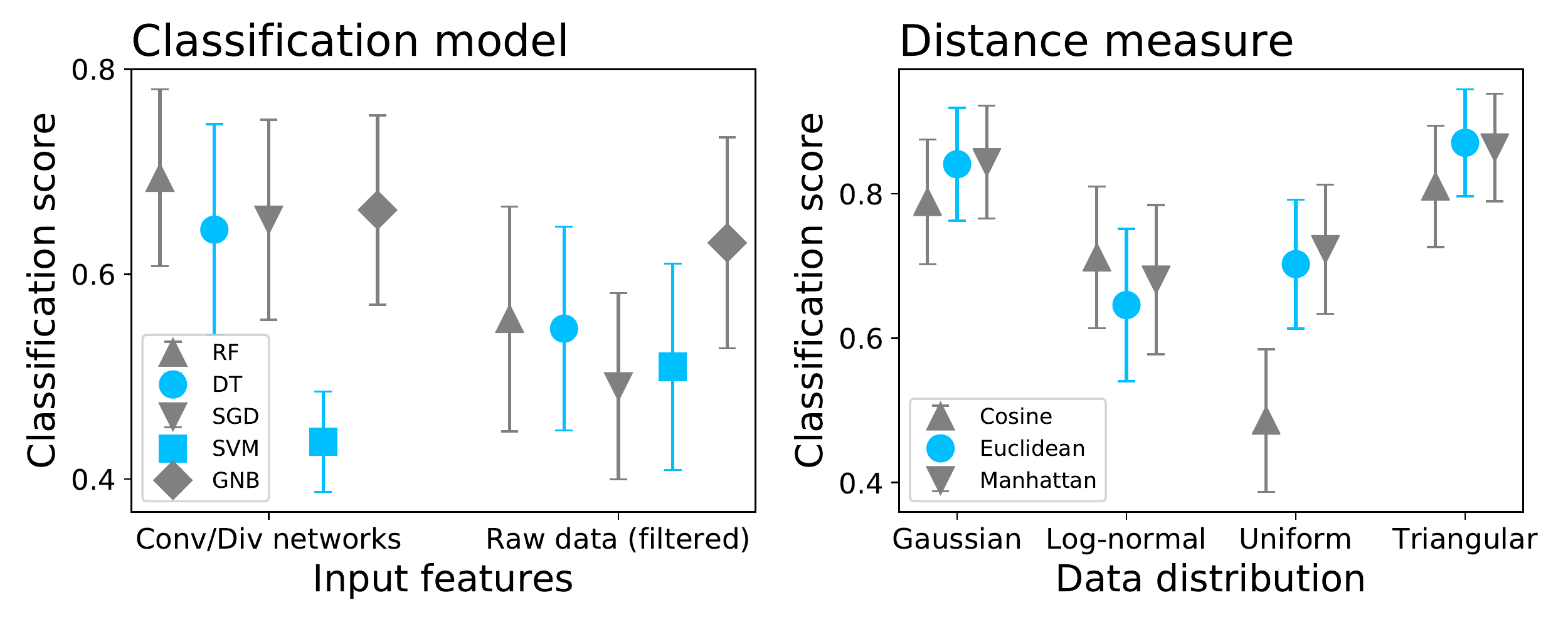}
\caption{Additional sensitivity analyses. (Left) Classification score, as obtained by five different classification algorithms with the convergence/divergence networks' (left group of symbols) and the raw data (right group). (Right) Classification score as a function of the underlying data probability distribution, and of the considered distance function. In both panels, symbols and whiskers respectively represent the average value and the corresponding standard deviation, as obtained in $10^4$ independent realisations. Parameters were fixed to $n_i = 50$, $\sigma_2 = 1.5$, $n_r = 5$ and $n_t = 15$.}\label{fig:ResOtherDM}
\end{center}
\end{figure}

As for the first, {\it i.e.} the possibility of using different classification algorithms beyond RF, Fig. \ref{fig:ResOtherDM} Left reports the average and the standard deviation of the classification score, as obtained with the convergence/divergence networks (left part) and the raw filtered data (right part) by several standard classification algorithms: Random Forests (RF) \cite{breiman2001random}, Decision Trees (DT) \cite{murthy1998automatic}, Stochastic Gradient Descent (SGD) with a hinge loss function \cite{ketkar2017stochastic}, Support Vector Machine (SVM) with radial kernels \cite{noble2006support}, and Gaussian Na\"ive Bayes (GNB).

Differences between the five algorithms are small, the Random Forest being the most efficient in both data sets. Nevertheless, two exceptions have to be highlighted.
Firstly, Support Vector Machines yield an extremely low classification score using the convergence/divergence networks. Being this result a clear outlier, it is most probably caused by some idiosyncrasies of the SVMs, than by the method here proposed.
Secondly, the Gaussian Na\"ive Bayes algorithm yields good results in both cases, {\it i.e.} even with the raw data alone. This is nevertheless due to the way the problem is designed ({\it i.e.} by sampling instances from normal distributions) and to the way this classification algorithm is constructed ({\it i.e.} by fitting data to normal distributions of unknown mean and variance). In other words, GNBs perfectly fit the hypothesis underlying the synthetic data, and this results in an increase in the average classification score. Yet, when this advantage is eliminated and instances are sampled from uniform distributions, {\it caeteris paribus}, the classification score in raw data reduces from $0.631$ to $0.583$, {\it i.e.} to the same level of the other algorithms.

Fig. \ref{fig:ResOtherDM} Right further reports the average classification score as obtained when three additional probability distributions were used to create the raw data: log-normal ($\mu = 0.0$, $\sigma = \sigma_2$), uniform and triangular (both in the range $[-\sigma_2, \sigma_2]$). Each symbol also represents a different algorithm for calculating the distance between instances: cosine, euclidean and manhattan. Results are qualitatively similar, with the log-normal and uniform distributions yielding the lowest classification scores - possibly due to the resulting distribution of outliers. Similarly, the use of a different distance measure does not substantially impact the output. The only significant exception arises for the combination uniform distribution / cosine distance, for which the algorithm fails to detect any pattern in the data; this effect seems to be robust, and does not depend on the type of classification model used.

\section{Validation with real biomedical data sets}
\label{sec:val:real}

The validation of the proposed methodology has further been performed using a collection of real biomedical data sets, thus allowing to understand if and when the convergence / divergence concept is of relevance in biology and medicine. For that, nine data sets corresponding to the ``Life science'' area and to a classification problem have been downloaded from the UCI Machine Learning Repository \cite{UCI}. Inclusion criteria were a limited number of instances, thus representing problems for which it is common to have small samples; and the availability of numerical features. The choice of using a public data set has been motivated by three considerations: first, the public availability of those data implies that all results can be seamlessly reproduced (all data sets can be freely downloaded from \url{https://archive.ics.uci.edu/ml/index.php}); second, the website provides references to many research works in which those data have been used, thus offering the possibility of comparing our results with relevant literature; and third, it allows to test the proposed methodology on multiple and diverse scenarios.

\begin{figure}[!tb]
\begin{center}
\includegraphics[width=0.99\textwidth]{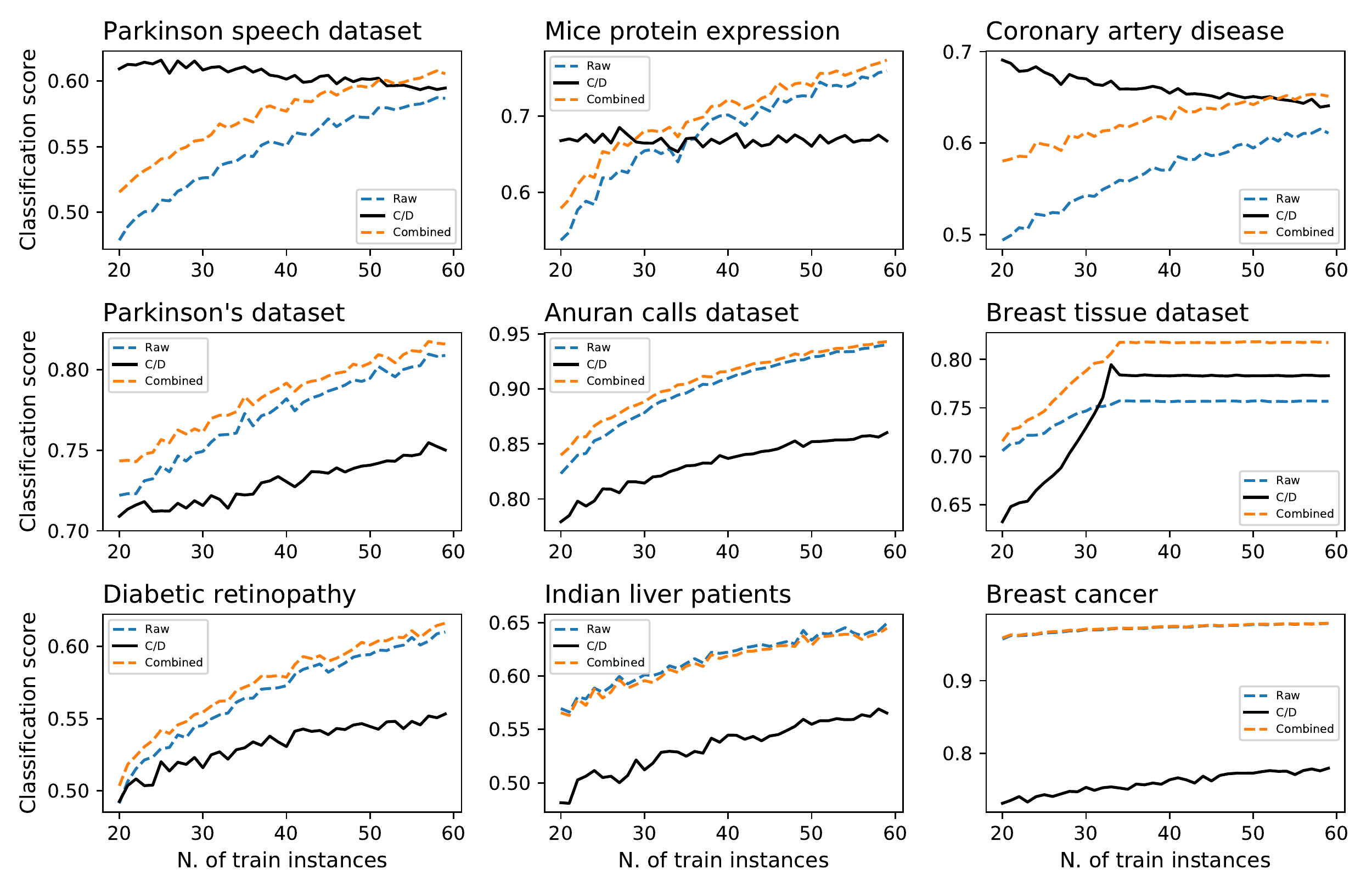}
\caption{Evolution of the classification score (in terms of the AUC) as a function of the number of subjects in the training set, and for the nine UCI data sets here considered - see main text and Tab. \ref{tab:ListFeatures} for details. The three lines respectively correspond to the score obtained with the raw features (blue dashed line), with the convergence / divergence features (solid black), and with both sets of features (dashed orange). Each point corresponds to the average AUC obtained in $2000$ random realisations.}\label{fig:ResOtherDS}
\end{center}
\end{figure}

Table \ref{tab:ListFeatures} lists the set of features used to create the convergence / divergence networks, along with references in which more details about the data sets can be found. Fig. \ref{fig:ResOtherDS} depicts the evolution of the average classification score for each data set, in terms of the resulting AUC, as a function of the number of instances used in the training, when such classification is performed with a Random Forest model \cite{breiman2001random} and a Leave-One-Out Cross Validation. Three scores are compared: {\it i}) the AUC obtained with the raw data set (blue dashed line); {\it ii}) the AUC for the classification performed using only the convergence / divergence metric (black solid line); and {\it iii}) the AUC obtained when combining both sets of instances (green dashed line). Note that the second metric quantifies how much information about the original problem is retained, and thus how effective are the convergence / divergence metrics in the dimensionality reduction; while the third how complementary the two metrics are, with respect to the raw features. Additional information for each data set, including histograms of the selected features and the distribution of the convergence / divergence metrics, are included in Supplemental Information.

It can be appreciated from Fig. \ref{fig:ResOtherDS} that the convergence / divergence metrics yield heterogeneous classification performances, mainly depending on the data set (and hence, on its characteristics) and on the number of instances.

First, it can happen that the classification score obtained with the convergence / divergence networks is above what obtained with the raw features - see the top row of Fig. \ref{fig:ResOtherDS}. This advantage is smaller for high number of instances, as standard classification algorithms are able to approximate the convergence/divergence pattern. Nevertheless, even for large training groups, the addition of the C/D metrics to the raw features improves the final classification.
Such scenario is clearly the most interesting one, as it implies that, on one hand, the proposed methodology is able to extract the most important information from the instance's characteristics, thus enabling an important dimensionality reduction. On the other hand, when very few instances are available, it is even able to extract information usually not unveiled.

In the second scenario, the proposed methodology is able to extract information that, while not especially useful in itself, can be used to improve a standard classification task. This is represented by the second row of data sets in Fig. \ref{fig:ResOtherDS}, for which the orange dashed line (representing the combined data set) is always above the blue one (raw features only). In these cases, the convergence / divergence mechanism here hypothesised is not the only mechanism underlying the pathological situation; yet, the networks help in making it more explicit, and thus complement a more standard approach.

Finally, it is possible that the proposed approach is not relevant for the problem under study: thus, both the C/D metrics, and their combination with the raw features yield results below those of a standard classification model - as can be observed in the last row of Fig. \ref{fig:ResOtherDS}. This is not surprising, as not all biomedical problems have to be explained by the mechanism here hypothesised. 

For the sake of completeness, the results corresponding to the first three data sets are analysed in more detail below, and compared with those available in the literature.

\begin{table*}[!tb]
\centering
\begin{tabular}{|p{5.2cm}|p{3.7cm}|p{3.7cm}|}
\hline
{\bf Data set} & {\bf Convergence features} & {\bf Divergence features} \\ \hline
\multirow{2}*{Parkinson Speech \cite{sakar2013collection}} & Local jitter & Absolute jitter \\
 & Median pitch & Shimmer (dda) \\ \hline
\multirow{2}*{Mice Protein Expression \cite{higuera2015self}} & pCAMKII & pELK \\
 & pNR2B & pNR1 \\ \hline
\multirow{2}*{Coronary Artery Disease \cite{alizadehsani2013data}} & Triglyceride & Creatine \\
 & Erithrocyte sedimentation rate & Triglyceride \\ \hline
\multirow{2}*{Parkinson's \cite{little2007exploiting}} & Max. fundamental frequency & Avg. fundamental frequency \\
 & DFA & Shimmer (APQ5) \\ \hline
\multirow{2}*{Anuran Calls \cite{colonna2016recognizing}} & MFCC$_2$ & MFCC$_1$ \\
 & MFCC$_4$ & MFCC$_4$ \\ \hline
\multirow{2}*{Breast Tissue \cite{jossinet1996variability}} & Phase angle & Impedance distance \\
 & Slope of p.a. & Length of s.c. \\ \hline
\multirow{2}*{Diabetic Retinopathy \cite{antal2014ensemble}} & MA$_1$ & Exudates$_3$ \\
 & Exudates$_1$ & Diameter \\ \hline
\multirow{2}*{Indian Liver Patient \cite{ramana2012critical}} & Direct bilirubin & Alamine Aminotransferase \\
 & Alkaline phosphotase & Aspartate Aminotransferase \\ \hline
\multirow{2}*{Breast Cancer Wisconsin \cite{street1993nuclear}} & Fractal dimension II & Fractal dimension I \\
 & Concavity III & Area II \\ \hline
\end{tabular}
\caption{List of the nine considered UCI data sets. The second and third columns respectively report the features used to create the convergence / divergence networks, selected according to the procedure described in Sec. \ref{sec:creation}.}
\label{tab:ListFeatures}
\end{table*}

\subsubsection*{Parkinson Speech Dataset.}
This data set was designed to support speech pattern analyses of Parkinson's patients and control subjects, with the final aim of providing a simple diagnostic tool. Multiple types of sounds, including sustained vowels, numbers, words and short sentences, were recorded for each subject, for then extracting $26$ amplitude, frequency and harmonicity based features \cite{sakar2013collection}.
Results, as depicted in Fig. \ref{fig:ResOtherDS} top left, indicate that the proposed method is especially suitable for analysing small sets of instances drawn from this data set.
Interestingly, one of the features, Shimmer (DDA), has previously yielded conflicting results, with evidence in favour \cite{jimenez1997acoustic} and against \cite{rahn2007phonatory} its relevance in diagnosing Parkinson's. Our results indicate that the average value is not relevant for the classification (as the corresponding histograms for control subjects and patients are very similar, see Supplemental Information); but is instead the intra-group variability, when combined with that of the absolute jitter, that is relevant.
While this data set is quite similar to the Parkinson's one \cite{little2007exploiting}, both in terms of scope and feature extracted, the benefit yielded by the proposed approach in the latter case is substantially smaller - see Fig. \ref{fig:ResOtherDS}. We speculate that this may be caused by the type of sounds recorded in each one of them - in the Parkinson's case, the only available information is that they were ``sustained vowel phonations'' \cite{little2009suitability}.
Finally, some additional results, as the evolution of the score as a function of the number of trees included in the Random Forest, and as a function of the number of features used to create the convergence / divergence networks, are included in Supplemental Information.

\subsubsection*{Mice Protein Expression Dataset.}
This second data set encodes the expression levels of $77$ proteins, measured in the cerebral cortex, for control and trisomic (Down syndrome's \cite{davisson1993segmental}) mice \cite{higuera2015self}. Note that the original division into he four groups, {\it i.e.} stimulated / not stimulated to learn, and injected with saline / memantine, has here been disregarded.
It is worth highlighting that three of the four identified proteins ({\it i.e.} pCAMKII, pELK and pNR1) were originally identified as relevant features for distinguish memantine-treated mice {\it vs.} control baseline\cite{higuera2015self}; and that levels of the two N-methyl-D-aspartate receptor subunits pNR2B and pNR1 have been found to be associated to improved learning following an electroacupuncture treatment \cite{lu2016electroacupuncture}.

\subsubsection*{Coronary Artery Disease Dataset.}

The third test case, known in the UCI repository as the Z-Alizadeh Sani Data Set, comprises information about control subjects and people diagnosed with a coronary artery disease, the latter one presenting a narrowing of the coronary artery of $50\%$ or more. Features include demographic, symptom, ECG, and laboratory results \cite{alizadehsani2013data}. 
Two important results have to be highlighted. First of all, none of the identified features were considered as important in the original work - only the erithrocyte sedimentation rate (ESR) ranked 11 over 54 in terms of importance in the final classification model. Indeed, if one observes the value distributions of, {\it e.g.}, creatine and triglyceride alone, these are very similar for both patients and control subjects; yet their combination unveil a higher variability in control subjects - see Supplemental Information.
Secondly, it is worth noting that triglyceride appears as a relevant feature both in the creation of the convergence and of the divergence networks - a topic which is further discussed in Supplemental Information.

\section{Discussion and conclusions}

In this contribution we presented a computational framework, based on complex networks, that allows to change the way symptoms and signs are defined. Instead of identifying patterns involving differences in the mean of a given feature between two (or more) groups, we here focus in changes in the variability. 
If the latter aspect is usually disregarded by classical classification algorithms, we here showed that it can be used to reconstruct a network structure, with nodes representing subjects, and links the pairwise similarity between them; and that such network structure can be used to improve the performance in a classification task.
This approach has firstly been tested with synthetic data sets, to understand its behaviour under a wide range of controlled conditions; and secondly, following the hypothesis that the feature variability should especially be relevant in biology and medicine, with real data sets representing a wide range of biomedical problems. In some cases a significant increase in the classification score has been observed, thus suggesting that the proposed mechanism is indeed of relevance.

Both Figs. \ref{fig:ResSynth} and \ref{fig:ResOtherDS} indicate that the proposed methodology is especially well suited for the analysis of small data sets - as similar results can be recovered by standard algorithms only in the limit of a very high number of instances. This may {\it prima facie} seem an important limitation, due to the increasing relevance of ``big data'' in today's scientific and clinical environments \cite{obermeyer2016predicting}. Nevertheless, the opposed movement is also gaining momentum: due to increasing privacy and security concerns, or to the need of stratifying instances to a much smaller scale, many applications have to be developed relying on very small data sets \cite{estrin2014small, kitchin2015small}. 

Additionally, both Fig. \ref{fig:ResOtherDS} and the discussion of Sec. \ref{sec:val:real} suggest that the proposed framework can be used, in some cases, as a way to identify novel biomarkers. Due to the way features are processed in the convergence / divergence networks, some of them may be useless when considered alone, but of relevance when interacting - as is the case of the features identified in the Coronary Artery Disease Dataset.

Finally, it is worth noting that, while we here focused on classification tasks, the conclusions' validity goes well beyond the data mining field. When the addition of a set of synthetic features ({\it i.e.} constructed on top of the available data) results in an increase in the classification score, it can safely be assumed that these new features are making explicit some information that was originally encoded in the data, but in a form not suitable to be digested by the data mining model. The increased score here observed confirms that the proposed approach is describing the biomedical problems in a way that is more efficient (or explicit) than what done by the raw data; or, in other words, that data are transformed from the raw features' to a more useful space.

\section*{Authors' contributions}
MZ conceived the study and participated in data and statistical analyses; JMT participated in the design of the study and in data analysis; EM coordinated the study. All authors contributed to the preparation of the manuscript, and gave final approval for publication.

\section*{Competing interests}
We have no competing interests.


\bibliographystyle{vancouver}

\clearpage

\section*{\Huge Supplemental Information}

\vspace{1cm}

\section{Methodology: supplemental analyses}

\subsection{Overlaps between the convergence and divergence sets}

As seen in the analysis of the real biomedical data sets included in the main text, it can happen that a same feature is recognised as simultaneously optimising the convergence and divergence networks. While being something counterintuitive, we are here going to show a possible mechanism explaining the appearance of such situations.

\begin{figure*}[!b]
\begin{center}
\includegraphics[width=0.99\textwidth]{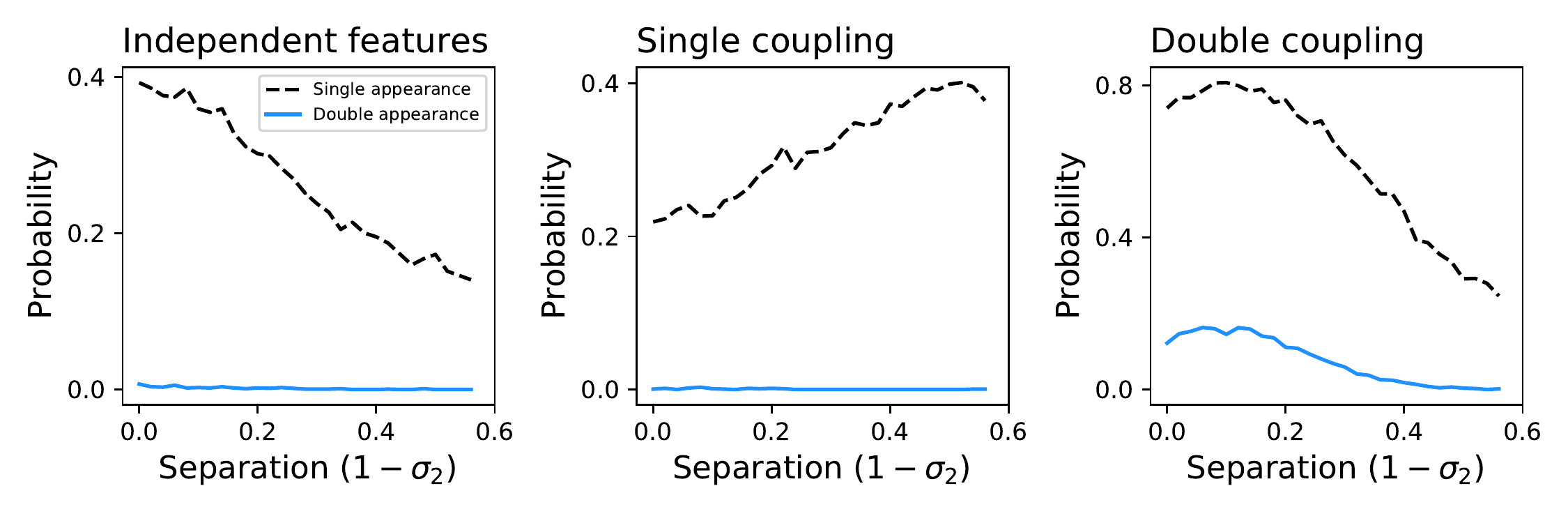}
\caption{Evolution of the probability of having an irrelevant feature selected during the creation of the convergence / divergence networks, as a function of the separation between the two groups. See the main text for details on the interpretation.}
\label{fig:DoubleFeature}
\end{center}
\end{figure*}

First of all, let us consider a synthetic data set structured as illustrated in the Tab. I of the main text. Specifically, and for the sake of simplicity, we will consider a situation with the first two features being relevant for the first group, two more being relevant for the second group, and six uncorrelated features - {\it i.e.} $n_r = 2$ and $n_t = 10$.
In an ideal situation, we would expect features $1$ and $2$ (respectively, $3$ and $4$) to be identified as relevant for the first group (for the second group). Nevertheless, when the increase in the standard deviation of the relevant features (denoted by $1 - \sigma_2$) is small enough, it may happen that an irrelevant feature is selected instead, just due to random fluctuations.
The probability of such event to happen is plotted in Fig. \ref{fig:DoubleFeature} Left, as a function of the standard deviation separation. Specifically, the black dashed line represents the probability of an irrelevant feature to be selected at least in one set; and the solid blue line the probability for an irrelevant feature to be selected to be part of both sets at the same time.
As may be expected, the probability of a single appearance of an irrelevant feature is maximal when $\sigma_2 = 1$, and decreases for $\sigma_2 > 1$; On the other hand, the probability of a double appearance is negligible.

As a second case, we consider the possibility of an irrelevant feature to be partly correlated with one of the relevant features. For instance, let us suppose the following case: $f_5 = f_1 + \mathcal{N}(0, 0.1 \sigma_2)$, $\mathcal{N}$ representing numbers independently drawn from a normal distribution. The evolution of the two probabilities, again as a function of the separation $1 - \sigma_2$, is represented in Fig. \ref{fig:DoubleFeature} Centre.
In this case, the probability of finding the irrelevant (now correlated) feature in one of the two sets increases with the separation - in other words, the feature is wrongly included in the set even for $\sigma_2 \gg 1$. 
In order to understand this, let us consider the case of two uncorrelated features $f$ and $g$, each one comprising values drawn from an uniform distribution $\mathcal{U}(-1, 1)$. Given two instances, defined by a pair $(f, g)$, the average distance between them is given by the formula $\frac{2}{15}( 2 + \sqrt{2} + 5 \ln (1 + \sqrt{2}) )$, which is $\approx 1.04$. If we now consider a new feature $h$ strongly correlated with $f$, and suppose that instances are defined by pairs $(f, h)$, the average distance between pairs of them is equivalent to the average distance between two points located on a segment of length $2 \sqrt{2}$, which yields $\frac{2}{3} \sqrt{2} \approx 0.94$.
In synthesis, the average distance between pairs of instances is on average smaller when considering features that are correlated; thus, the uncorrelated feature is selected more often than what expected, as it increases the coherence of the convergence set.

Finally, let us consider a third situation, in which one irrelevant feature is correlated with two relevant features belonging to different groups. Specifically, one can define $f_5$ such that $f_5(i) = f_1(i) + \mathcal{N}(0, 0.1 \sigma_2)$ for instances $i$ belonging to the first group, and $f_5(i) = f_3(i) + \mathcal{N}(0, 0.1 \sigma_2)$ otherwise. In other words, the feature $f_5$, which was initially irrelevant, is now correlated with a relevant feature of both groups of instances. This has two effects - see Fig. \ref{fig:DoubleFeature} Right. First, it strongly increases the probability for $f_5$ of appear in at least one group; but it further make the probability of being part of both groups of features non-negligible.

\clearpage

\section{Validation with real biomedical data sets: details}
\label{sec:RealBiomed}

This section includes detailed information on the convergence/divergence analyses presented in Section {\it Validation with real biomedical data sets} of the main paper. Specifically, for each one of the nine data sets considered, the following images report:

\begin{itemize}

	\item {\bf Convergence set and metrics.} Information on the creation of the convergence network, including (from left to right):
	\begin{itemize}
		\item Scatter plot of the two features for which the first set of instances is more similar than the second one.
		\item Histograms of the values of the two features, divided according to the instances' group.
		\item Histogram of the final convergence metric.
	\end{itemize}
	
	\item {\bf Divergence set and metrics.} Information on the creation of the divergence network, including (from left to right):
	\begin{itemize}
		\item Scatter plot of the two features for which the first set of instances is more heterogeneous than the second one.
		\item Histograms of the values of the two features, divided according to the instances' group.
		\item Histogram of the final divergence metric.
	\end{itemize}
	
	\item {\bf Final convergence network.} For the sake of clarity, only the nodes belonging to the giant component of the network are displayed, for a maximum of 100 randomly chosen nodes (when available). Additionally, low weight links have been pruned to keep an average degree of 5.

	\item {\bf Final divergence network.} The same rules as for the representation of the convergence network apply.

\end{itemize}

\begin{figure*}[!tb]
\begin{center}
\includegraphics[width=0.99\textwidth]{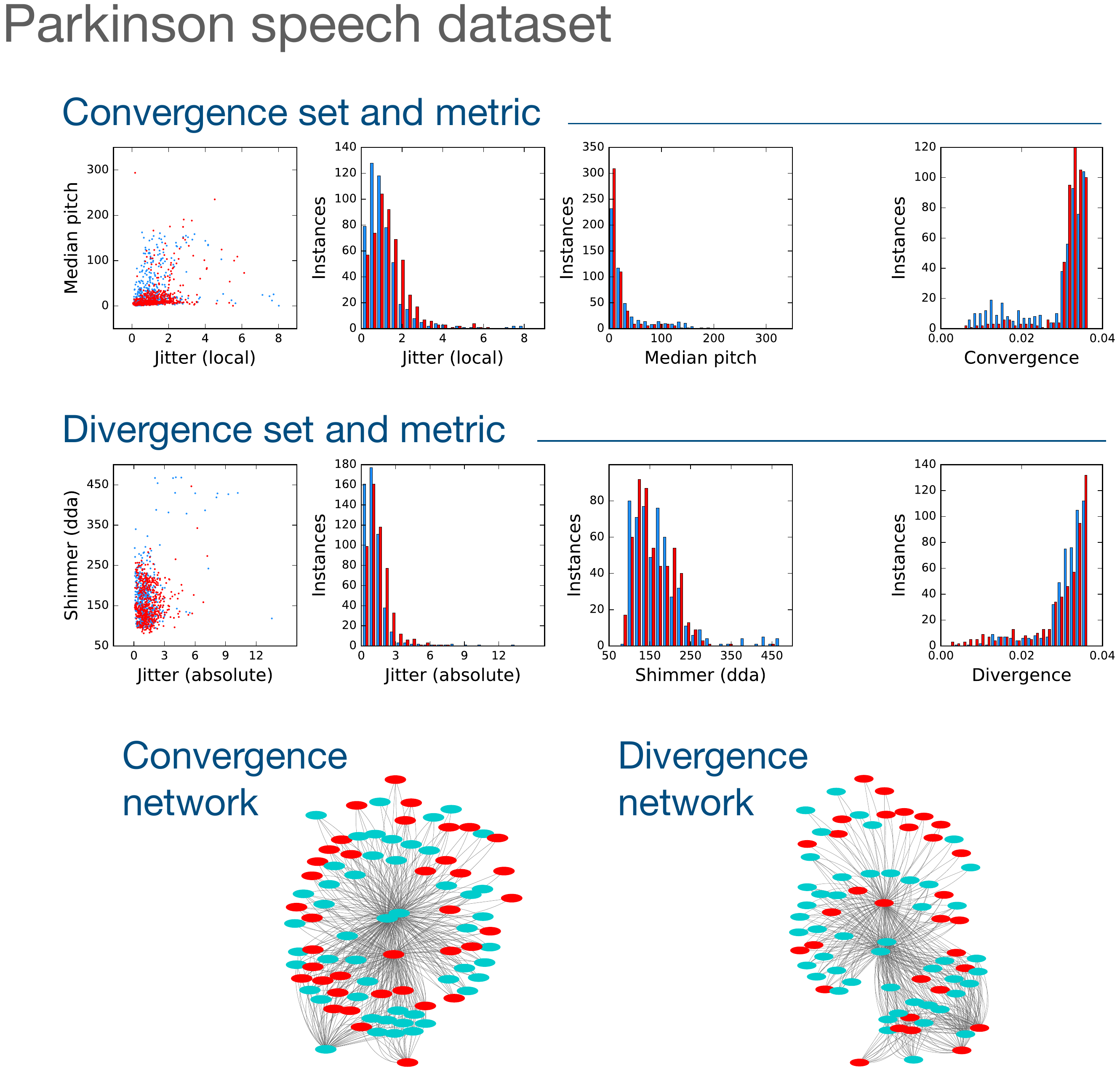}
\label{fig:DS1}
\end{center}
\end{figure*}

\begin{figure*}[!tb]
\begin{center}
\includegraphics[width=0.99\textwidth]{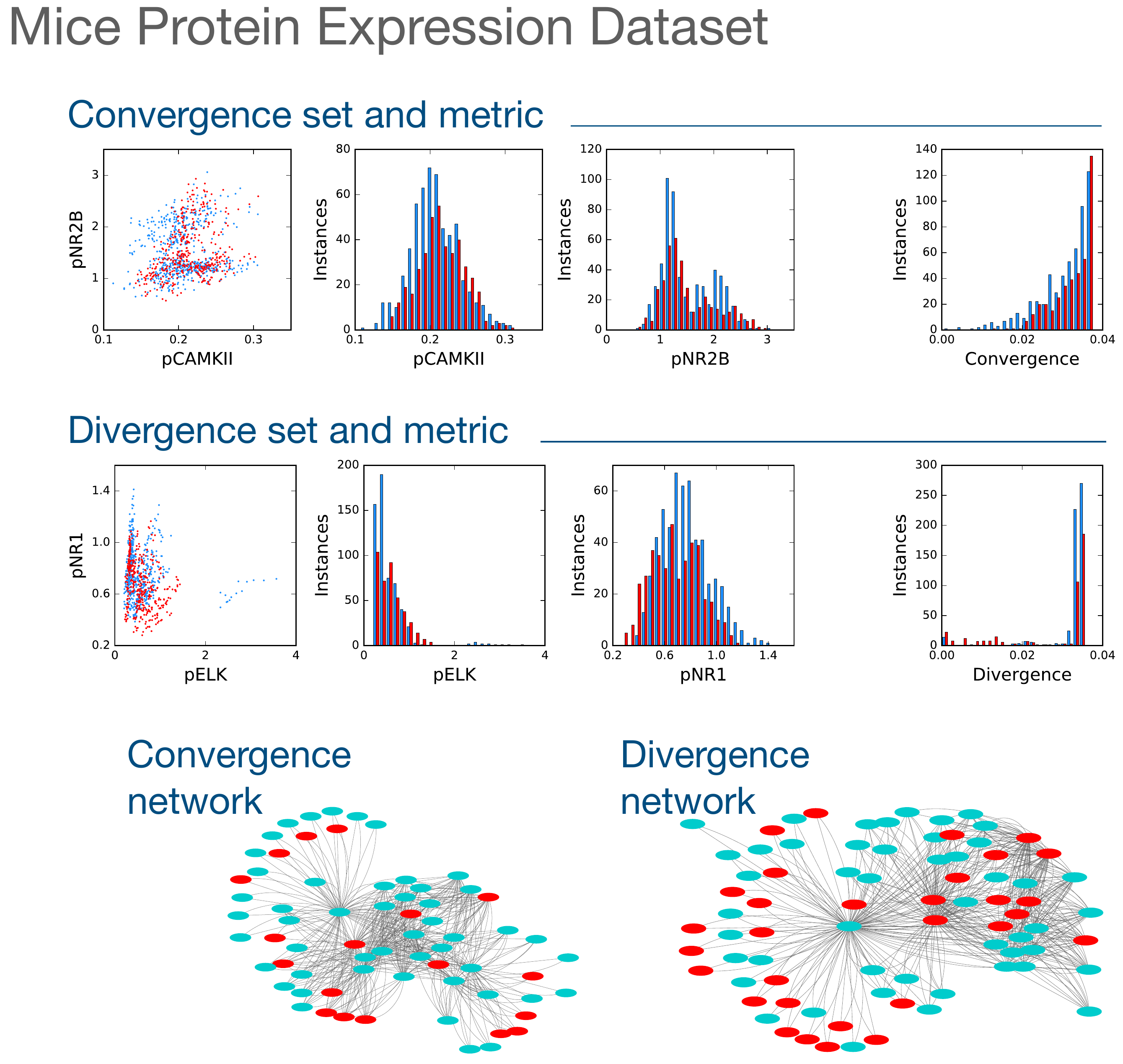}
\label{fig:DS2}
\end{center}
\end{figure*}

\begin{figure*}[!tb]
\begin{center}
\includegraphics[width=0.99\textwidth]{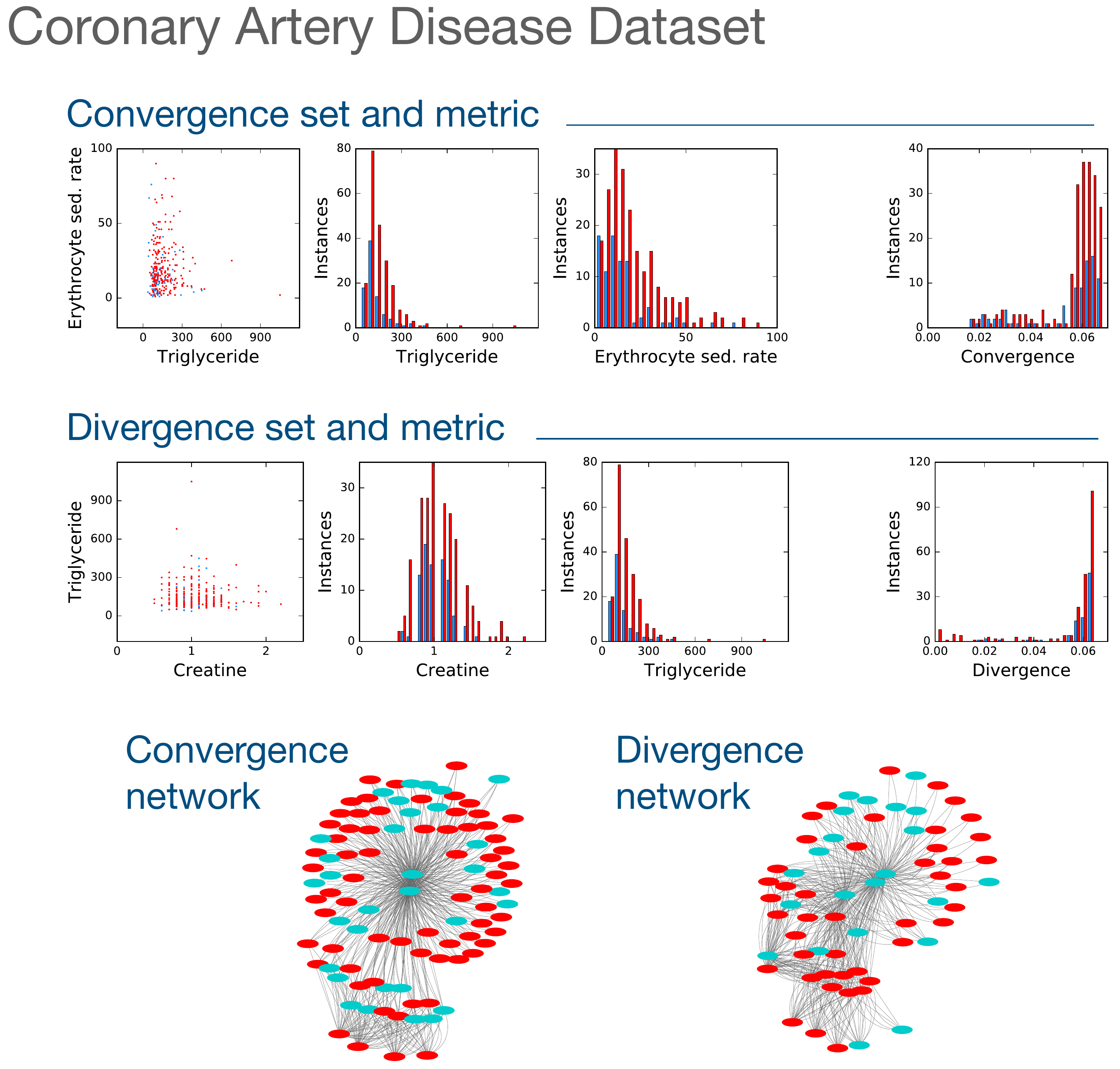}
\label{fig:DS3}
\end{center}
\end{figure*}

\begin{figure*}[!tb]
\begin{center}
\includegraphics[width=0.99\textwidth]{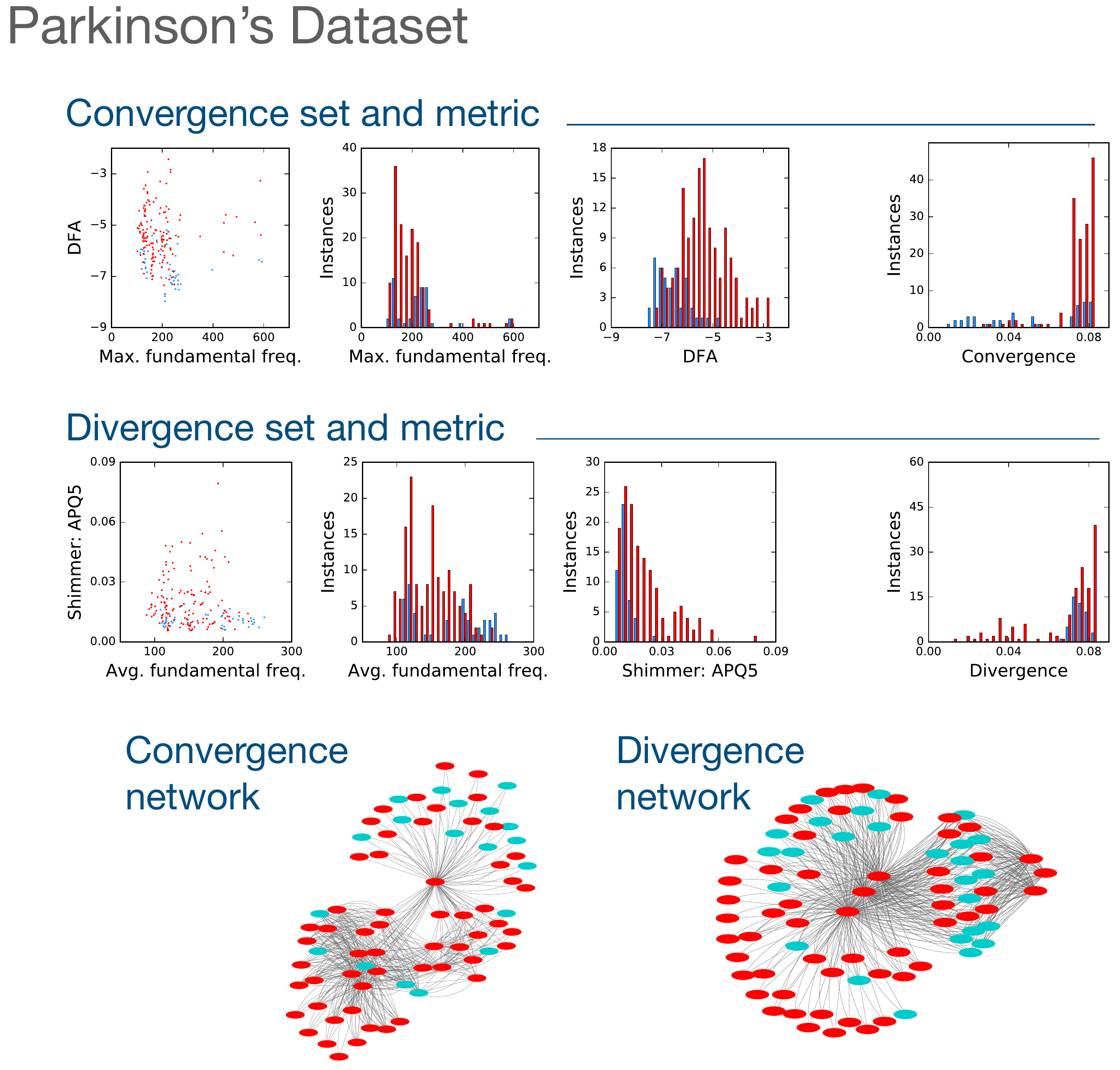}
\label{fig:DS4}
\end{center}
\end{figure*}

\begin{figure*}[!tb]
\begin{center}
\includegraphics[width=0.99\textwidth]{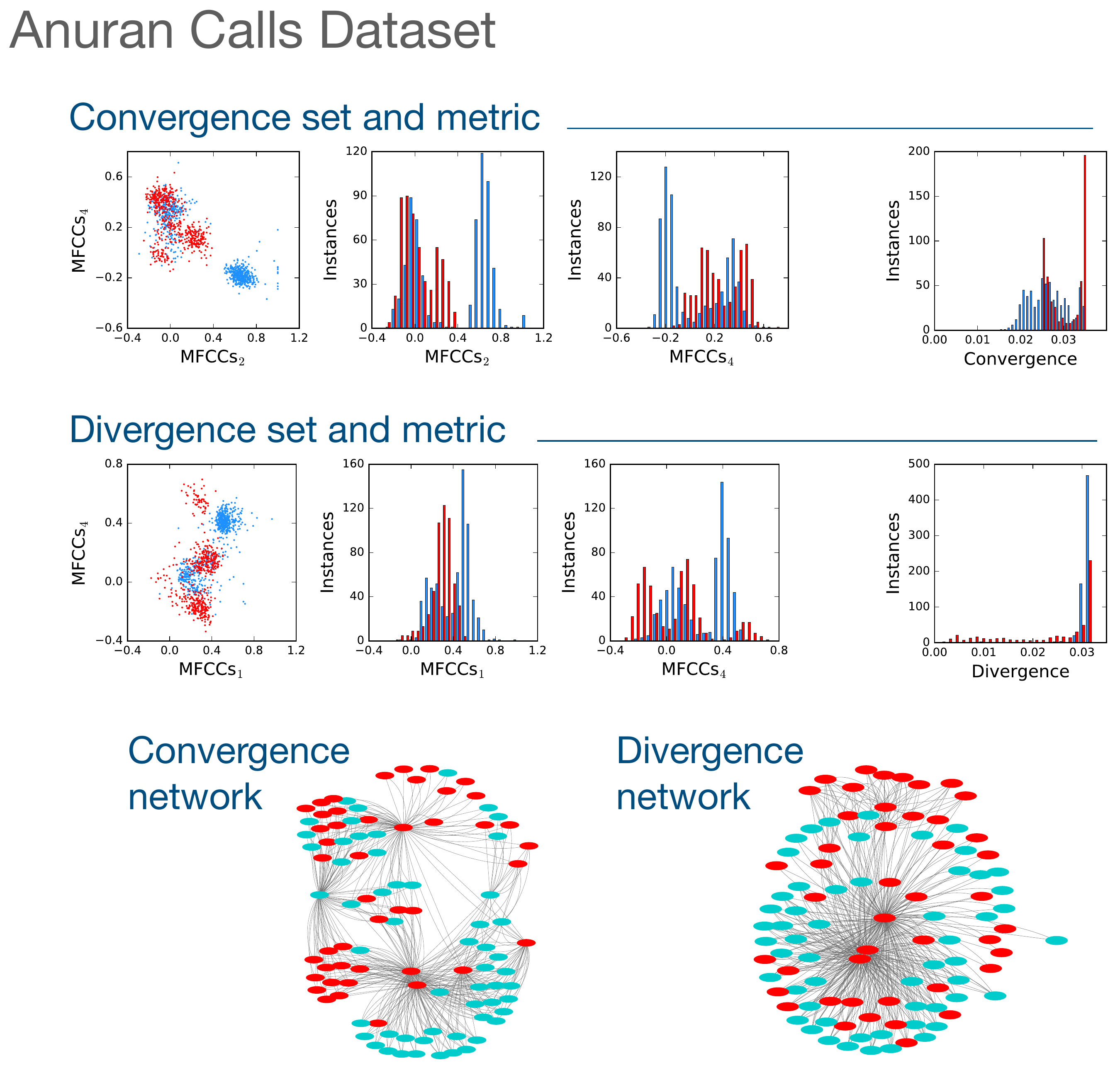}
\label{fig:DS5}
\end{center}
\end{figure*}

\begin{figure*}[!tb]
\begin{center}
\includegraphics[width=0.99\textwidth]{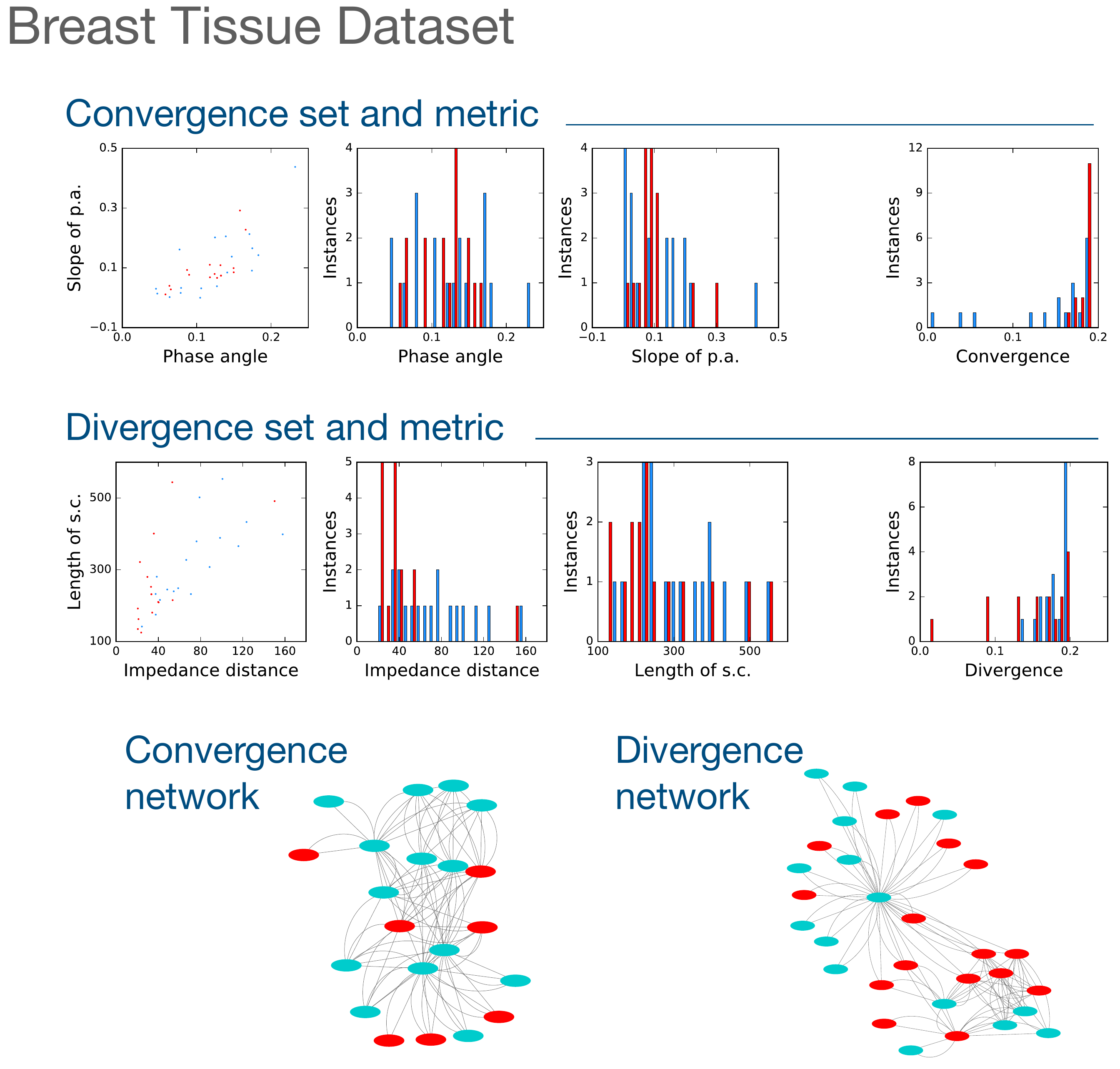}
\label{fig:DS6}
\end{center}
\end{figure*}

\begin{figure*}[!tb]
\begin{center}
\includegraphics[width=0.99\textwidth]{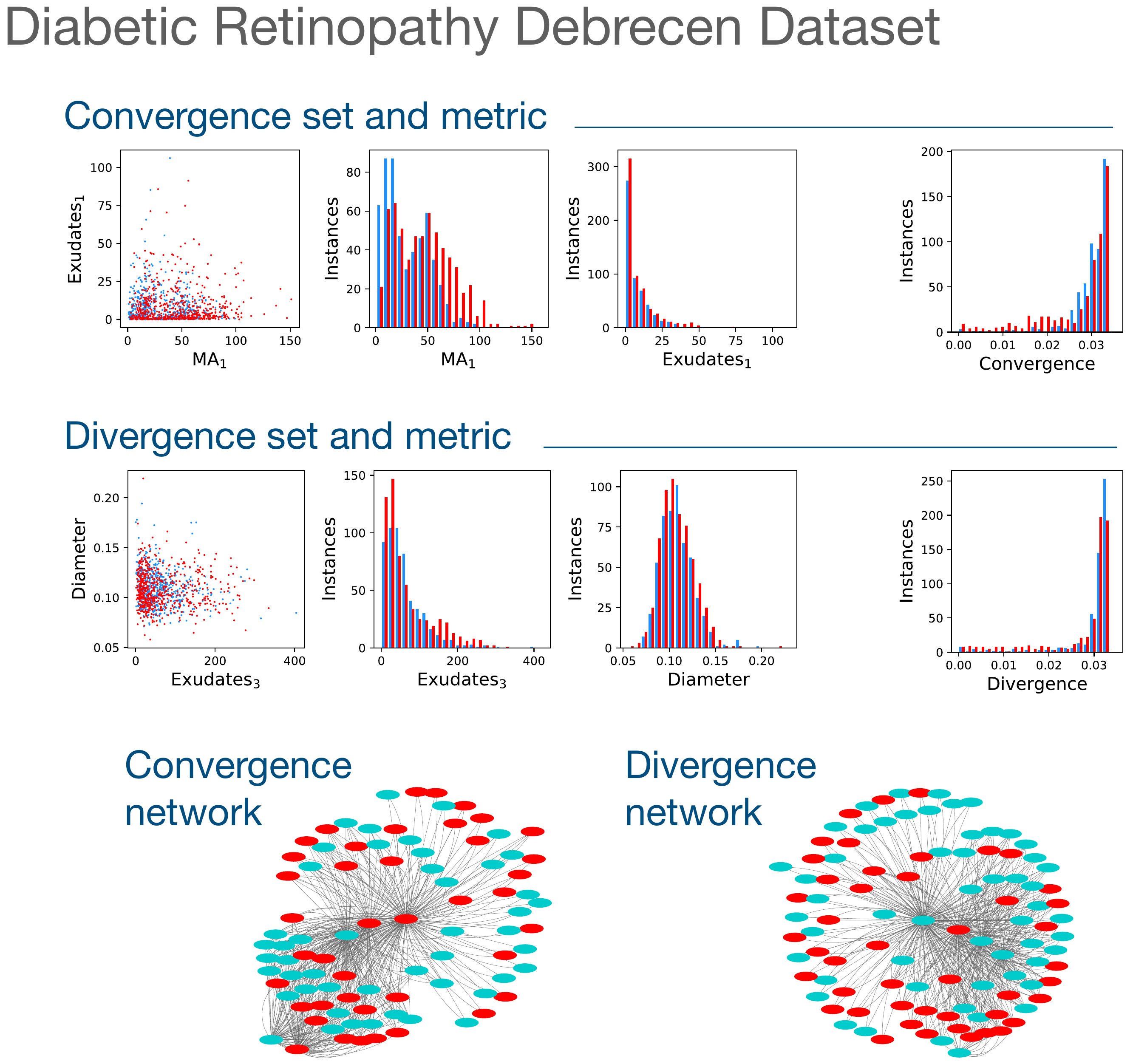}
\label{fig:DS7}
\end{center}
\end{figure*}

\begin{figure*}[!tb]
\begin{center}
\includegraphics[width=0.99\textwidth]{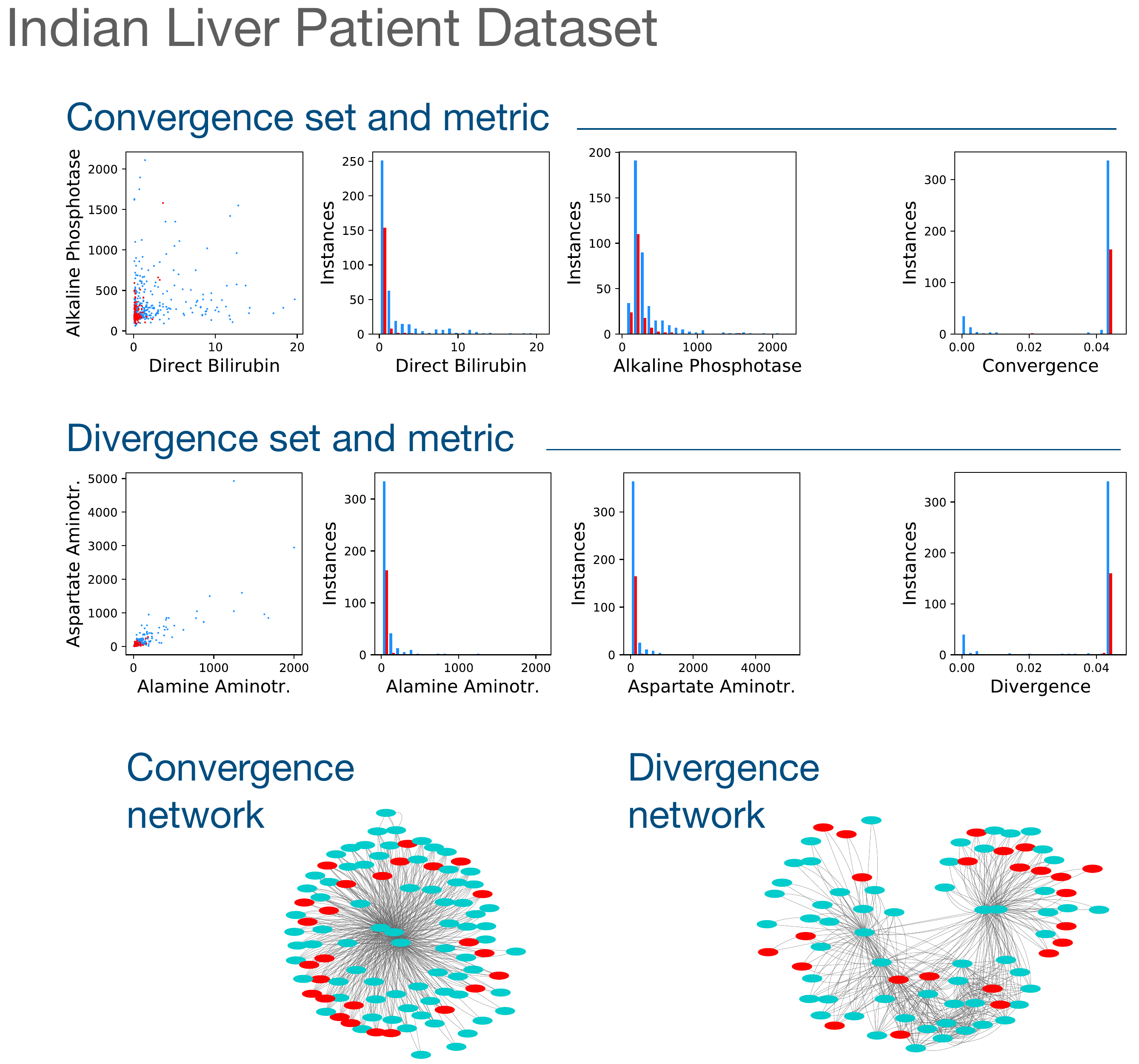}
\label{fig:DS8}
\end{center}
\end{figure*}

\begin{figure*}[!tb]
\begin{center}
\includegraphics[width=0.99\textwidth]{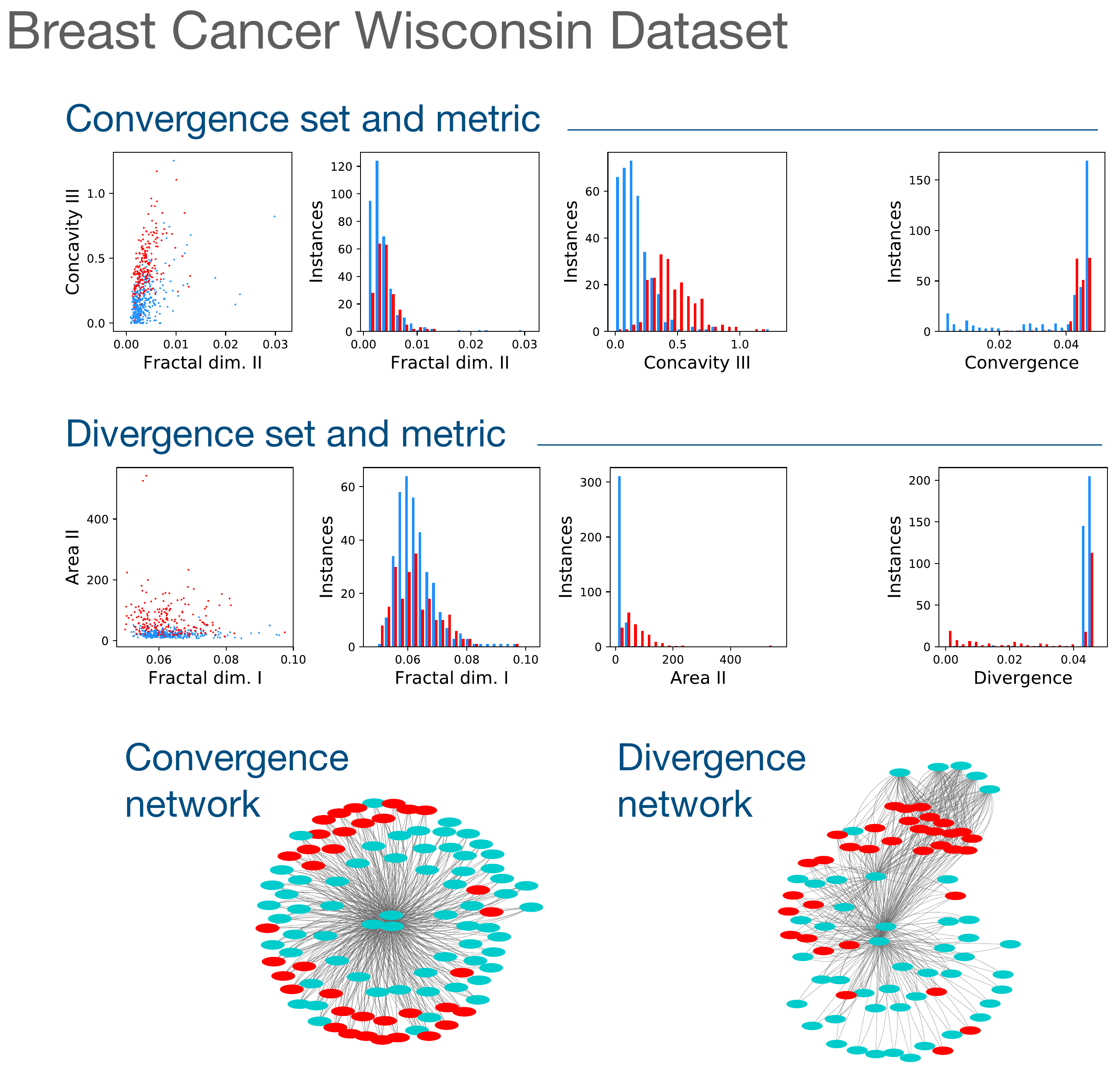}
\label{fig:DS9}
\end{center}
\end{figure*}

\clearpage

\section{Additional classification analyses}
\label{sec:AddAnalyses}

In this section we further report some results related with the classification task executed on the Parkinson's speech dataset.

\begin{figure*}[!h]
\begin{center}
\includegraphics[width=0.99\textwidth]{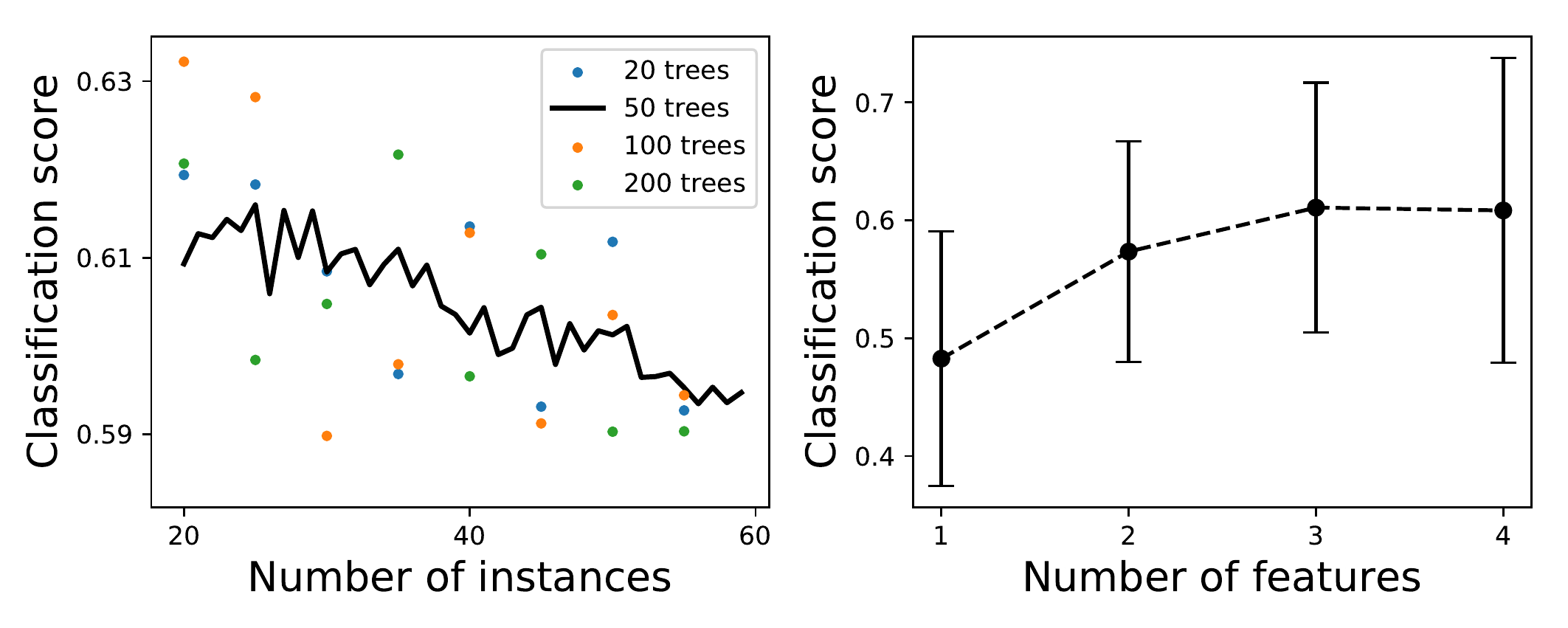}
\caption{(Left) Evolution of the classification score in the Parkinson's speech dataset, as a function of the number of instances included in the training, and for different numbers of trees trained in the Random Forest. (Right) Evolution of the classification score, for the same data set, as a function of the number of features selected to create the convergence / divergence networks.}
\label{fig:AddAnalyses}
\end{center}
\end{figure*}

Fig. \ref{fig:AddAnalyses}, Left panel, reports the evolution of the classification score as a function of the number of instances used in the training, using the two centralities extracted from the convergence / divergence networks. Additionally to the information reported in Fig. 4 of the main text, we report the score obtained with Random Forests comprising 20, 100 and 200 trees. It can be appreciated that the results are qualitatively similar, as, due to the limited number of features used for training, there is little risk of overfitting.

Fig. \ref{fig:AddAnalyses}, Right panel, depicts the classification score (with 30 instances for training, and 50 trees) as a function of the number of features used in the creation of the convergence / divergence networks. Results suggest that including more than two features does not substantially improve the efficacy of the task, while it importantly increase the computational cost.

\end{document}